\documentclass[namedreferences]{kluwer}
\usepackage[]{epsfig}
\def\lsim{\vcenter{\hbox{$<$}\offinterlineskip\hbox{$\sim$}}}
\def\gsim{\vcenter{\hbox{$>$}\offinterlineskip\hbox{$\sim$}}}
\begin{document}
\begin{article}

\begin{opening}
\title{Structure and evolution of the inner Milky Way galaxy}
\subtitle{results from ISOGAL}
\author{Jacco \surname{van Loon}\email{jacco@ast.cam.ac.uk}}
\institute{Institute of Astronomy, Madingley Road, Cambridge CB3 0HA, United
Kingdom}
\author{for The ISOGAL Collaboration}
\institute{}
\runningtitle{The Structure and Evolution of the inner Milky Way Galaxy}
\runningauthor{Jacco van Loon}
\begin{abstract}

The DENIS/ISOGAL near/mid-IR survey of the Milky Way for the first time probes
stellar populations in the innermost obscured regions of our galaxy. Ages,
metallicities and extinction-corrected luminosities are derived for these
stars individually. An old metal-rich population dominates in the inner
galactic Bulge, but there are also indications for the presence of a younger
population. The inner Bulge has a tri-axial shape, as traced by depth effects
on the observed luminosity distributions.

\end{abstract}
\end{opening}

\section{Introduction}

Attempts to understand the evolution of galaxies by studies of their stellar
populations as a function of redshift are limited by sensitivity and angular
resolution. Our nearest galaxy, the Milky Way, has the potential to provide
many important clues on the evolution of galaxies: ages and metallicities may
be measured for individual stars, and the spatial and kinematic distributions
of the different stellar populations may be observed. For the innermost parts
of the Milky Way, however, where most stellar light and mass are and where
most activity is happening, this has not yet been the case. Due to the
location of the Sun in the galactic plane, our view of the central few hundred
pc of the galactic Bulge and the inner few kpc of the galactic disk is
obscured by tens of magnitudes extinction at visual wavelengths.

Advances in IR instrumentation have recently led to several near- and mid-IR
surveys of the Milky Way, both from the ground (DENIS, 2MASS, TMGS) and from
space (ISOGAL, MSX). These probe large numbers of stars situated deep within
the obscured regions of the inner Bulge, for the first time allowing the study
of the stellar populations in the core of our own galaxy. The catalogue of IR
point sources from the DENIS/ISOGAL 0.8---15 $\mu$m survey \cite{O99} is used
here to derive the ages, metallicities, luminosities and extinction for
$\sim3\times10^4$ individual stars in the inner galactic Bulge, and the
implications for the structure and evolution of the galactic Bulge are
discussed.

\section{Data and methods}

\begin{figure}
\centerline{\includegraphics[width=100mm]{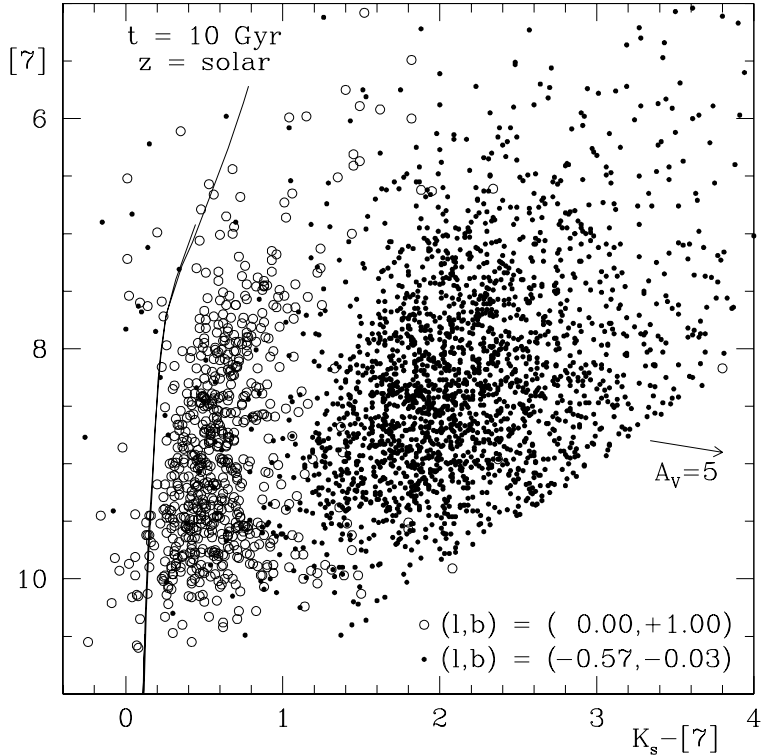}}
\caption{The $[7]$ versus $(K-[7])$ diagram of the stellar populations in two
fields in the inner Galaxy that suffer from different amounts of (severe)
extinction. An isochrone for a 10 Gyr population of solar metallicity is
plotted for illustration.}
\end{figure}

The ISOGAL 7 and 15 $\mu$m images of typically $15^\prime\times15^\prime$ and
pixel scales of $3^{\prime\prime}$ or $6^{\prime\prime}$ sample the Milky Way
especially well for $-10^\circ<l_{\rm II}<+10^\circ$ and $-2^\circ{\lsim}\
b_{\rm II}\ {\lsim}+2^\circ$, which is the region studied here. Photometry for
point sources is complemented by DENIS I,J and K$_{\rm s}$-band photometry. An
example of an IR colour-magnitude diagram for two fields is given in Fig.\ 1,
together with an isochrone for a 10 Gyr old population of solar metallicity
\cite{B94}. The red colours of the stars are predominantly due to severe
interstellar extinction.

The age, metallicity and extinction (adopting \opencite{M90}) are derived for
each individual star by comparing its location with respect to isochrones
\cite{B94} in various IR colour-magnitude diagrams, after proper computation
of bolometric corrections. For stars without sufficient photometry to solve
for all three variables, the extinction is assigned as derived for
neighbouring stars. The luminosity and effective temperature are obtained for
free.

\section{Stellar populations}

\begin{figure}
\centerline{\includegraphics[width=100mm]{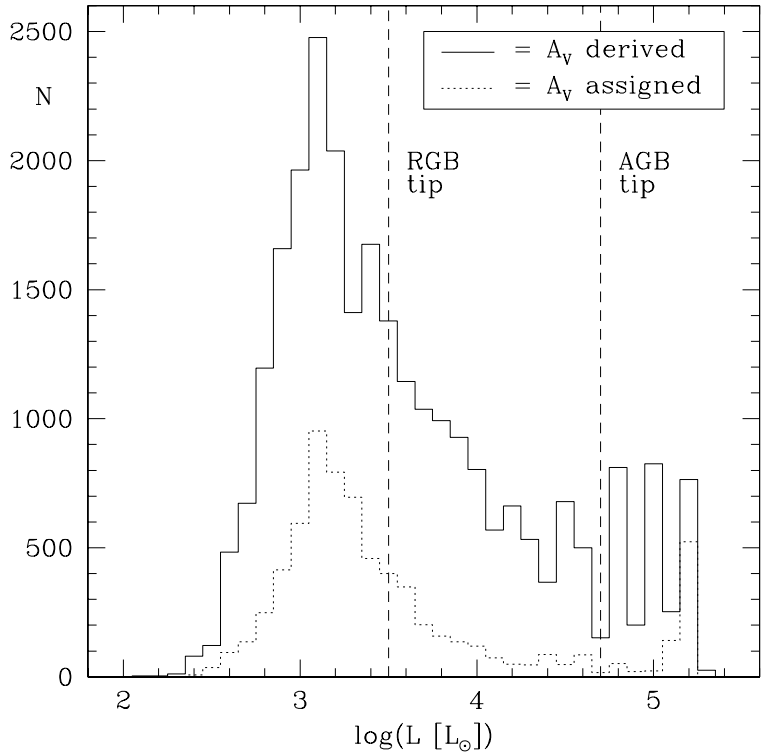}}
\caption{Luminosity distribution derived from the comparison of IR photometry
and isochrones. All of the AGB, and a significant portion of the RGB are
detected.}
\end{figure}

As a first result, the luminosity distribution of the stars is plotted in
Fig.\ 2. The Asymptotic Giant Branch (AGB) is detected even in the obscured
galactic core region. In most regions the Red Giant Branch (RGB) is detected
down to a few $10^2$ L$_\odot$. The derived effective temperatures are
generally $T_{\rm eff}\sim2500$ to 4000 K, confirming the red giant nature of
most of the stars. A few hundred bright mid-IR sources have been identified
that are interpreted as mass-losing AGB stars, some of which are associated
with OH maser emission.

\section{Ages and metallicities}

The metallicity distribution covers the wide range of $[M/H]{\in}[-1.7,0.4]$,
although the higher metallicity stars amongst these are more common. There is
only a very marginal indication for a negative metallicity gradient over the
inner $\sim1$ kpc from the Galactic Centre.

The dominant population of the inner Bulge is old, $t{\gsim}10$ Gyr, but there
is a significant intermediate population, $t\sim1$ Gyr, and possibly an even
younger population of $t{\lsim}100$ Myr too. The age distribution is fairly
uniform over the inner Bulge, although the average age of the youngest
component may slightly increase outwards.

This suggests that in the galactic Bulge star formation did not switch off
completely after the first generations of stars were born and that we now see
as the old, metal-rich population. Instead, the intermediate-age population of
AGB stars implies on-going star formation over most of the galactic history,
while the young population might represent a recent epoch of enhanced star
formation, possibly due to a minor merger event.

\section{Three-dimensional structure of the inner galactic Bulge}

The extinction-corrected luminosity function shows a clear asymmetry in the
galactic plane at either side of the Galactic Centre: it is brighter at
$l_{\rm II}\sim-6^\circ$ than at $l_{\rm II}\sim+6^\circ$. This can be
understood in terms of differential depth effects, if the inner Bulge has a
tri-axial shape on a radial scale of $R\sim1.4$ kpc under an angle
$i\sim53^\circ$, with the near side at negative longitude. This geometry is
much alike that proposed for the larger scale Bar, which is, however, rather a
disk phenomenon. The origin of the tri-axiality of the Bulge might be related
to the suggested recent star formation.

\begin{acknowledgements}
I thank the organisation for giving me the opportunity to present this work at
a pleasant and interesting conference. Special thanks to Sne\v{z}ana for being
there and Joana for being everywhere.
\end{acknowledgements}

{}

\end{article}
\end{document}